\input harvmac
\input epsf
\noblackbox
\newcount\figno
\figno=0
\def\fig#1#2#3{
\par\begingroup\parindent=0pt\leftskip=1cm\rightskip=1cm\parindent=0pt
\baselineskip=11pt
\global\advance\figno by 1
\midinsert
\epsfxsize=#3
\centerline{\epsfbox{#2}}
\vskip 12pt
\centerline{{\bf Figure \the\figno} #1}\par
\endinsert\endgroup\par}
\def\figlabel#1{\xdef#1{\the\figno}}
\def\pano{\par\noindent}

\def\pmb#1{\setbox0=\hbox{#1}%
 \kern-.025em\copy0\kern-\wd0
 \kern.05em\copy0\kern-\wd0
 \kern-.025em\raise.0433em\box0 }
\font\cmss=cmss10
\font\cmsss=cmss10 at 7pt

\def\rlx{\relax\leavevmode}
\def\Cop{\relax\,\hbox{$\kern-.3em{\rm {\bf C}}$}}
\def\Rop{\relax{\rm I\kern-.18em R}}
\def\Nop{\relax{\rm I\kern-.18em N}}
\def\Pop{\relax{\rm I\kern-.18em P}}
\def\Zop{\rlx\leavevmode\ifmmode\mathchoice{\hbox{\cmss Z\kern-.4em Z}}
 {\hbox{\cmss Z\kern-.4em Z}}{\lower.9pt\hbox{\cmsss Z\kern-.36em Z}}
 {\lower1.2pt\hbox{\cmsss Z\kern-.36em Z}}\else{\cmss Z\kern-.4em
 Z}\fi}


\def\H{{\cal H}}

\def\ie{{\it i.e.}}


\lref\polvtwo{J. Polchinski, {\it String Theory, Volume 2, Superstring
Theory and Beyond}, CUP 1998.}

\lref\vafawitten{C. Vafa, E. Witten, {\it On orbifolds with discrete
torsion}, J. Geom. Phys. {\bf 15}, 189 (1995); {\tt hep-th/9409188}.}

\lref\stef{B. Stefa\'nski, {\it Dirichlet Branes on a Calabi-Yau 
three-fold orbifold}, Nucl. Phys. {\bf B589}, 292 (2000);
{\tt hep-th/0005153}.}

\lref\vafa{C. Vafa, {\it Modular invariance and discrete torsion on
orbifolds}, Nucl. Phys. {\bf B273}, 592 (1986).}

\lref\gomis{J. Gomis, {\it D-branes on orbifolds with discrete torsion
and topological obstruction}, JHEP {\bf 0005}, 006 (2000); 
{\tt hep-th/0001200}.}

\lref\douglas{M.R. Douglas, {\it D-branes and discrete torsion}, {\tt
hep-th/9807235}.}

\lref\dougfiol{M.R. Douglas, B. Fiol, {\it  D-branes and discrete
torsion II}, {\tt hep-th/9903031}.}

\lref\mrgrev{M.R. Gaberdiel, {\it Lectures on Non-BPS Dirichlet
branes}, Class. Quant. Grav. {\bf 17}, 3483 (2000);
{\tt hep-th/0005029}.}

\lref\gabste{M.R. Gaberdiel, B. Stefa\'nski, {\it Dirichlet
branes on orbifolds}, Nucl. Phys. {\bf B578}, 58 (2000); 
{\tt hep-th/9910109}.} 

\lref\bgtwo{O. Bergman, M.R. Gaberdiel, {\it Stable non-BPS
D-particles}, Phys. Lett.~{\bf B441}, 133 (1998);
{\tt hep-th/9806155}.}

\lref\mukray{S. Mukhopadhyay, K. Ray, {\it D-branes on fourfolds with
discrete torsion},  Nucl. Phys. {\bf B576}, 152 (2000); 
{\tt hep-th/9909107}.}

\lref\berlei{D. Berenstein, R.G. Leigh, {\it Discrete torsion, AdS/CFT
and duality}, JHEP {\bf 0001}, 038 (2000); {\tt hep-th/0001055}.}

\lref\diagom{D.-E. Diaconescu, J. Gomis, {\it Fractional branes and
boundary states in orbifold theories}, {\tt hep-th/9906242}.}

\lref\kleinraba{M. Klein, R. Rabadan, {\it Orientifolds with discrete
torsion}, JHEP {\bf 0007}, 040 (2000); {\tt hep-th/0002103}.}

\lref\dasguptaetal{K. Dasgupta, S. Hyun, K. Oh, R. Tatar, {\it
Conifolds with Discrete Torsion and Noncommutativity}, JHEP 
{\bf 0009}, 043 (2000); {\tt hep-th/0008091}.}

\lref\kleinrabaone{M. Klein, R. Rabadan, {\it $Z_N \times Z_M$
orientifolds with and without discrete torsion}, JHEP {\bf 0010}, 049
(2000); {\tt hep-th/0008173}.}

\lref\aspinpless{P.S. Aspinwall, M.R. Plesser, {\it D-branes, Discrete
Torsion and the McKay Correspondence}, {\tt hep-th/0009042}.}

\lref\aspinwall{P.S. Aspinwall, {\it A Note on the Equivalence of
Vafa's and Douglas's Picture of Discrete Torsion}, JHEP {\bf 0012},
029 (2000); {\tt hep-th/0009045}.}

\lref\berleigh{D. Berenstein, R.G. Leigh, {\it Non-Commutative
Calabi-Yau Manifolds}, {\tt hep-th/0009209}.}

\lref\hankol{A. Hanany, B. Kol, {\it  On orientifolds, discrete
torsion, branes and M theory}, JHEP {\bf 0006}, 013 (2000);
{\tt hep-th/0003025}.}

\lref\aspmor{P.S. Aspinwall, D.R. Morrison, {\it Stable singularities
in string theory}, Commun. Math. Phys. {\bf 178}, 115 (1996);
{\tt hep-th/9503208}.}

\lref\sharpe{E.R. Sharpe, {\it Discrete torsion and gerbes I \& II},
{\tt hep-th/9909108} and {\tt 9909120}.}

\lref\sharpetwo{E.R. Sharpe, {\it Discrete torsion},
{\tt hep-th/0008154}.}

\lref\aadds{C. Angelantonj, I. Antoniadis, G. D'Appollonio, E. Dudas,
A. Sagnotti, {\it Type I vacua with brane supersymmetry breaking}, 
Nucl. Phys. {\bf B572}, 36 (2000); {\tt hep-th/9911081}.}

\lref\dougmoore{M.R. Douglas, G. Moore, {\it D-branes, quivers and ALE
instantons}, {\tt hep-th/9603167}.}

\lref\gab{M.R. Gaberdiel, {\it Discrete torsion orbifolds and
D-branes}, JHEP {\bf 0011}, 026 (2000); {\tt hep-th/0008230}.}

\lref\bcf{M. Bill\'o, B. Craps, F. Roose, {\it Orbifold boundary states
from Cardy's condition}, {\tt hep-th/0011060}.}

\lref\gimpol{E.G. Gimon, J. Polchinski, {\it Consistency conditions
for orientifolds and D-manifolds}, Phys. Rev. {\bf D54}, 1667 (1996);
{\tt hep-th/9601038}.}

\lref\doughull{M.R. Douglas, C. Hull, {\it D-branes and the
non-commutative torus}, JHEP {\bf 9802}, 008 (1998);
{\tt hep-th/9711165}.} 

\lref\hananyetal{B. Feng, A. Hanany, Y.-H. He, N. Prezas, 
{\it Discrete Torsion, Non-Abelian Orbifolds and the Schur
Multiplier}, {\tt hep-th/0010023} and {\it Discrete Torsion, Covering
Groups and Quiver Diagrams}, {\tt hep-th/0011192}.}

\lref\bjl{D. Berenstein, V. Jejjala, R.G. Leigh, {\it D-branes on
Singularities: New Quivers from Old}, {\tt hep-th/0012050}.}

\lref\wittenK{E. Witten, {\it D-branes and K-theory}, JHEP
{\bf 9812}, 019 (1998); {\tt hep-th/9810188}.}

\lref\bdl{M. Berkooz, M.R. Douglas, R.G. Leigh, {\it Branes
intersecting at angles}, Nucl. Phys. {\bf B480}, 265 (1996);
{\tt hep-th/9606139}.}

\lref\gutperle{M. Gutperle, {\it Non-BPS D-branes and enhanced
 symmetry in an asymmetric orbifold}, JHEP {\bf 0008}, 036 (2000);
{\tt hep-th/0007126}.}

\lref\majsen{J. Majumder, A. Sen, {\it Non-BPS D-branes on a
Calabi-Yau Orbifold}, JHEP {\bf 0009}, 047 (2000); 
{\tt hep-th/0007158}.}

\lref\pap{G. Papadopoulos, {\it private communication}.}

\Title{\vbox{
\hbox{hep--th/0101143}
\hbox{EFI-2001-03}
\hbox{KCL-MTH-01-02}}}
{\vbox{\centerline{Discrete torsion orbifolds and D-branes II}
}}
\centerline{Ben Craps\footnote{$^\star$}{{\tt
e-mail: craps@theory.uchicago.edu}}$^{,a}$ and 
Matthias R.\ Gaberdiel\footnote{$^\dagger$}{{\tt
e-mail: mrg@mth.kcl.ac.uk}}$^{,b}$}
\bigskip
\centerline{\it $^a$Enrico Fermi Institute, University of Chicago}
\centerline{\it 5640 S. Ellis Av., Chicago, IL 60637, USA}
\centerline{\it $^b$Department of Mathematics, King's College London}
\centerline{\it Strand, London WC2R 2LS, U.K.}
\smallskip
\vskip2cm
\centerline{\bf Abstract}
\bigskip
\noindent The consistency of the orbifold action on open strings
between D-branes in orbifold theories with and without discrete
torsion is analysed carefully. For the example of the
$\,\Cop^3/\Zop_2\times\Zop_2$ theory, it is found that the consistency
of the orbifold action requires that the D-brane spectrum contains
branes that give rise to a conventional representation of the orbifold
group as well as branes for which the representation is projective. It
is also shown how the results generalise to the orbifolds
$\,\Cop^3/\Zop_N\times\Zop_N$, for which a number of novel features
arise. In particular, the $N>2$ theories with minimal discrete torsion
have non-BPS branes charged under twisted R-R potentials that couple
to none of the  (known) BPS branes. 
\bigskip

\Date{01/2001}

\newsec{Introduction}

One of the ways in which Dirichlet branes have played an important
r\^ole in string theory is that they enable us to obtain insight into
the background geometry by analysing the low-energy theory
(and in particular its moduli space) of a Dirichlet brane probe. One
class of theories for which this is of particular interest are
orbifolds with discrete torsion \vafa\ whose geometric interpretation
is only partially understood
\refs{\vafawitten,\aspmor,\sharpe,\sharpetwo}. The issue of
understanding D-branes for this class of theories has attracted some
interest recently 
\refs{\douglas,\dougfiol,\diagom,\mukray,\aadds,\berlei,\gomis, 
\kleinraba,\hankol,\dasguptaetal,\kleinrabaone,\aspinpless,
\aspinwall,\berleigh,\hananyetal,\bcf,\bjl}. 

A framework for describing D-branes for general orbifold theories was
developed in \refs{\gimpol,\dougmoore}. In this approach, one begins
with an invariant configuration of D-branes on the covering space and
restricts the open string spectrum to those states that are invariant
under the action of the orbifold group. This `total' action on the
open string states $|\psi,ij\rangle$ can be decomposed into an action
on the oscillator state $\psi$ and an action on the Chan-Paton factors
$ij$ 
\eqn\totalintro{
g |\psi,ij\rangle  
= \gamma(g)_{ii'} |U(g)\psi,i'j'\rangle\, \gamma(g)^{-1}_{j'j}\,.}
It was argued in \gimpol\ that the consistency of the group action
requires that $\gamma$ should be a conventional or a projective
representation of the orbifold group. 

For the case of orbifolds with discrete torsion, Douglas proposed 
\refs{\douglas} that D-branes are characterised by the property that 
the representation $\gamma$ that appears in \totalintro\ is a 
{\it projective} representation of the orbifold group. For the
simplest example where we consider the compactification on a torus
with a B-field (which induces torsion), this can be intuitively
understood as follows. In the presence of a non-trivial B-field, 
the world-volume theory of a Dirichlet brane is non-commutative
\refs{\doughull}, and this translates into a `non-commutative' 
({\it i.e.} projective) action of the orbifold group  
on the Chan-Paton factors of the open string.

In general, however, it was argued in \gab\ that the relation
between `discrete torsion' and `projective' representations of the
orbifold group is more involved. In particular, the specific example
of the $\Cop^3/\Zop_2\times\Zop_2$ orbifold with and without discrete 
torsion was analysed, and the relevant Dirichlet branes were
constructed using the boundary state approach. It was found that for
branes that are localised at the fixed point, the above relation
between discrete torsion and projective representations was
satisfied. However, both theories also have branes that carry the
other representation (\ie\ a projective representation for the theory
without discrete torsion, and a conventional representation for the
theory with discrete torsion). This was also shown to be necessary in
order for the D-brane spectrum to be invariant under T-duality (that
relates the theory with and without discrete torsion \vafawitten).    

The resulting D-brane spectrum, however, raises a
puzzle\footnote{$^\star$}{We thank Greg Moore for drawing our
attention to this problem.} that we shall resolve in this paper: since
each of the two theories has conventional as well as projective
Dirichlet branes, the representation $\gamma$ that appears in
\totalintro\ is a direct sum of a conventional and a projective
representation, and therefore neither conventional nor
projective. Thus the D-brane spectrum that was found in \gab\ appears
to be in conflict with the results of \gimpol. As we shall explain in
some detail, a careful application of the consistency analysis of
\gimpol\ actually implies that the theory must have both conventional
and projective Dirichlet branes in order to be consistent. This is a
consequence of the fact that the action of the orbifold group on the
oscillator states is  {\it not} a conventional representation for all
open string sectors (as was implicitly assumed in \gimpol), but rather
defines a projective representation for some open string sectors, and
a conventional representation for the others.\footnote{$^\dagger$}{It
was noted in \dougmoore, in the context of orientifold theories, that
this subtlety may occur.} 

We shall also analyse the non-BPS D-branes for both theories, paying
particular attention to the question of whether the representation of
the orbifold group on the Chan-Paton indices is conventional or
projective, and how this ties in with the above consistency analysis. 
\smallskip

The $\Cop^3/\Zop_2\times\Zop_2$ orbifold is in many ways special, and
it is {\it a priori} not clear if and in which way the above findings
generalise to more general orbifolds. In order to address this issue,
we also analyse the case of $\Cop^3/\Zop_N\times\Zop_N$ orbifolds
without discrete torsion and with minimal discrete torsion in quite
some detail. As we shall see, the analysis depends crucially on the
value of $N$, in particular on whether $N$ is odd, twice an odd
number, or divisible by four.

One remarkable result is that for $N>2$ the theory with minimal
discrete torsion has non-BPS D-branes that seem to be
stable against the decay into BPS brane anti-brane pairs
throughout the moduli space. These D-branes carry R-R charges that are
not carried by any of the BPS branes of the theory (that have
been considered before in \refs{\dougfiol,\diagom}). We also find that
for $N$ divisible by four, the theory without discrete torsion 
and the theory with minimal discrete torsion both
have fractional D-branes for which the orbifold action on the
Chan-Paton factors cannot even be written as a direct sum of
conventional and projective representations. These branes are
nevertheless (presumably) consistent since the action on the open
string oscillator states has the same property. 
\smallskip

Throughout the paper, we shall use the boundary state approach for the  
description and analysis of Dirichlet branes on orbifolds. We shall
briefly review some background material in the next subsection, and
refer the reader to \refs{\bgtwo,\diagom,\gabste,\mrgrev,\bcf} for
more details. We shall also briefly summarise some basic facts about
discrete torsion; a good introduction can be found in \gomis\ (see
also \refs{\gab,\bcf}).  
\medskip

The paper is organised as follows. The next subsection contains a
brief review of discrete torsion and D-branes on orbifolds.
In Section~2 we revisit the $\Cop^3/\Zop_2\times \Zop_2$ orbifold with
and without discrete torsion. We analyse the consistency of the branes
proposed in \refs{\gab}, thereby generalising the framework of
\gimpol. We also discuss some of the non-BPS D-branes in these
theories and analyse their consistency. In Section~3 we collect some
basic facts about the $\Cop^3/\Zop_N\times \Zop_N$ orbifolds that
shall form the centre of attention for the rest of the paper. The
D-brane spectrum of these theories is analysed for odd $N$ in
Section~4, and for even $N$ in Section~5. Section~6 contains some
conclusions and open questions.  

\subsec{Some facts about D-branes on orbifolds and discrete torsion}

For our purposes, an orbifold can be thought of as the quotient of a
manifold  by a discrete group. If the action of the discrete group on
the manifold is not free, \ie\ if some group elements have fixed
points, then the resulting space is singular. An example is the
quotient of the real plane $\Rop^2$ by the $\Zop_N$ subgroup of 
rotations around the origin. In this case the resulting space is a
cone with a curvature singularity at the origin. Despite such
classical singularities, string theory is well-behaved on
orbifolds.    

In order to describe in more detail the orbifold construction in
string theory, let us consider the example of a closed string theory 
with background ${\cal M}$ on which an (abelian) group $G$
acts as a group of symmetries. The orbifold theory by $G$
consists of those states in the original space of states that are
invariant under the action of the orbifold group $G$. In
addition, the theory has so-called twisted sectors containing strings 
that are closed in ${\cal M}/G$ but not in
${\cal M}$. If the orbifold action has singularities, the twisted
sector states describe degrees of freedom that are localised at the
singularities; the presence of these additional states is the
essential reason for why string theory is well-behaved despite these 
singularities.  

In the abelian case, there is one twisted sector $\H_h$ for each
element $h\in G$. Each twisted sector has to be projected again
onto the states that are invariant under the orbifold group $G$;
the corresponding projector is of the form  
\eqn\proj{ P = {1 \over |G|} \sum_{g\in G} g \,,}
and the total partition function of the theory is then 
\eqn\part{ Z 
= {1 \over |G|}  \sum_{g,h\in G} Z(g,h)\,,}
where 
\eqn\Zdef{Z(g,h) = 
\hbox{Tr}_{\H_h}(q^{L_0} \bar{q}^{\bar{L}_0} g) \,.}
{}From a conformal field theory point of view, the presence of the
twisted sectors is required by the condition that the total partition
function should be modular invariant. However, as was pointed out by
Vafa \refs{\vafa}, for certain orbifold groups
this condition does not uniquely determine the
resulting partition function. Indeed, if \part\ is modular invariant,
then so is 
\eqn\partfnDT{
Z={1\over |G|}\sum_{g,h\in G} \epsilon(g,h) Z(g,h)\,,
}
provided that the phases $\epsilon(g,h)$ satisfy 
\eqn\consfirst{\eqalign{ 
\epsilon(h_1 h_2,g) & = \epsilon(h_1,g) \epsilon(h_2,g)  \cr
\epsilon(h,g) & = \epsilon(g,h)^{-1}\,. }}
The relevant phases $\epsilon(g,h)$ are called discrete torsion
phases. The ambiguity that is described by these phases corresponds to
an ambiguity in the definition of the orbifold action in each twisted
sector. 

We now turn to the description of D-branes on orbifolds. For
concreteness, let us assume that spacetime is the product of Minkowski
space and an orbifold. We first consider a `bulk' brane that may be
extended  along some of the directions transverse to the orbifold but
that is localised at a generic point in the orbifold. The dynamics of
such a D-brane is described in terms of open  strings as
follows. Consider a preimage of the brane on the covering space and
add image branes to obtain a configuration invariant under the
orbifold  group. (Since we are considering a brane at a generic point
of the orbifold, we shall need a total of $|G|$ copies.)  Then
consider all open strings with endpoints on any (two) of these
branes. The excitations of the D-brane are described by the
open string states that are invariant under the action of
$G$. As mentioned before, the action of a group element on the
open string states can be written in terms of an action on the
Chan-Paton indices and an action on the oscillator states (see
\totalintro). In the case of a bulk brane the representation on the
Chan-Paton indices $\gamma$ is the regular representation of $G$
(which has indeed dimension $|G|$).

If the D-brane is localised at a singular point of the orbifold,
the dimension of $\gamma$ may be smaller. This is a consequence of
the fact that we need fewer preimages in the covering space to make an 
orbifold invariant configuration. Branes for which the dimension of
$\gamma$ is strictly smaller than $|G|$  are called `fractional'
D-branes. Because the dimension of $\gamma$ is smaller than that of a
bulk D-brane, fractional D-branes cannot move off the singular point;
instead, a number of fractional D-branes have to come together in
order for the system to be able to move off into the bulk. 

D-branes describe open string sectors that can be added consistently
to a given closed string theory. In order to analyse this consistency
condition, it is often useful to consider an annulus (or cylinder)
diagram for which the boundary conditions are determined by two
(possibly identical) D-branes --- one for each boundary. In the
simplest case we have a diagram without an insertion of a vertex
operator. This diagram can be given two different Hamiltonian 
interpretations, depending on which world-sheet coordinate is chosen
as the world-sheet time. From an open string point of view, the
diagram is interpreted as a one-loop vacuum diagram. In an orbifold
theory, this diagram will always contain a projector \proj\ that 
ensures that only orbifold invariant open string states run in the
loop. On the other hand, from the closed string point of view the
diagram describes the tree-level exchange of a closed string between
two sources (D-branes). Each D-brane can then be described by a
`boundary state' $|D\rangle$, a coherent state in the closed string
theory that describes the emission and absorption of closed string
states by the D-brane. The condition that both the open and the closed
string interpretations of the annulus diagram should be sensible
imposes strong restrictions on the  possible D-branes in a given
closed string theory.  

Let us close this brief review by summarising some of the most
important features of the boundary state construction for orbifold
theories. (More details can be found in
\refs{\bgtwo,\diagom,\gabste,\mrgrev,\bcf}.) In an orbifold theory a
boundary state is typically a sum of components that are defined
in each untwisted and twisted sector of the theory. The component in a
given sector describes the coupling of the D-brane to closed strings
in that sector. Bulk branes are described by boundary states whose only
non-vanishing components are in the untwisted sectors. On the other hand,  
the boundary state of a fractional brane has generically a non-trivial  
component in at least one twisted sector. The contribution of the
$h$-twisted sector to the cylinder diagram considered before
corresponds, from the open string point of view, to the one-loop
diagram with the insertion of $h$. The sum over twisted sectors
reproduces then the projection operator \proj. In particular, boundary
states with a non-trivial component in the $h$-twisted sector lead to
open strings for which the annulus diagram gets a non-trivial
contribution from the insertion of $h$. This implies that the
world-volume of the corresponding D-brane must intersect its image
under the action of $h$.

\newsec{The $\Zop_2\times \Zop_2$ case.}

Let us begin by reviewing the case of the $\Cop^3/\Zop_2\times\Zop_2$
orbifold with and without discrete torsion. The following discussion
extends the results and the consistency analysis of \gab. For
simplicity we shall consider the uncompactified theory. 

The orbifold group is generated by $g_1$ and $g_2$ where
$g_1^2=g_2^2=1$ and $g_1 g_2=g_2 g_1$. These generators act by
inversion on some of the coordinates $x^3,\ldots,x^8$. More
specifically, $g_1$ maps $x^i\mapsto -x^i$ for $i=5,6,7,8$ and $g_2$
maps $x^j\mapsto -x^j$ for $j=3,4,7,8$.  

The second cohomology group $H^2(\Zop_2\times\Zop_2,U(1))$ is
$\Zop_2$, and there are therefore two orbifold theories: the theory
without discrete torsion and the theory with discrete torsion, for
which $g_i$ acts in the $g_j$-twisted sector (where $i\ne j$) with a
relative minus sign. In order to describe the D-brane spectrum of
these theories, it is convenient to introduce the following notation:
we denote, as in \gabste, a Dirichlet $p$-brane by $(r;s_1,s_2,s_3)$
where $p=r+s_1+s_2+s_3$, provided that it has $r+1$ Neumann boundary
conditions along the directions that have not been affected by the
orbifold, \ie\ $x^0,x^1,x^2,x^9$, and $s_i$ Neumann boundary
conditions along the directions $x^{2i+1}$ and $x^{2i+2}$. We shall
always fix $x^0$ and $x^9$ to be the light-cone coordinates; $x^1$ and
$x^2$ are unaffected by the orbifold.

In the following we shall describe both type IIA and type IIB in a
uniform fashion. Most of the analysis will be the same for both cases,
the only difference being the possible values of $r$ for a given
choice of $s_i$. Unless specified otherwise, we will always assume
that $p=r+s_1+s_2+s_3$ is even in IIA and odd in IIB. 

It is well known that D-branes couple to R-R potentials. It is
therefore worthwhile to summarise the spectrum of R-R ground states of 
the orbifolds we are studying by giving their Hodge diamonds. In the
theory without discrete torsion, the untwisted sector contributes
\refs{\vafawitten} 
\eqn\hodge{
\matrix {& & & &1& & &  \cr
           & & &0& &0& &  \cr
           & &0& &3& &0&  \cr
           &1& &3& &3& &1 \cr
           & &0& &3& &0&  \cr
           & & &0& &0& &  \cr
           & & & &1& & &  } 
}
to the Hodge diamond. The total contribution of the three twisted
sectors is 
\eqn\hodgetw{
\matrix {& & & &0& & &  \cr
           & & &0& &0& &  \cr
           & &0& &3& &0&  \cr
           &0& &0& &0& &0 \cr
           & &0& &3& &0&  \cr
           & & &0& &0& &  \cr
           & & & &0& & & }  
~.}
In the theory with discrete torsion, the untwisted contribution is the
same, while the twisted sectors now contribute
\eqn\hodgetwdt{
\matrix {& & & &0& & &  \cr
           & & &0& &0& &  \cr
           & &0& &0& &0&  \cr
           &0& &3& &3& &0 \cr
           & &0& &0& &0&  \cr
           & & &0& &0& &  \cr
           & & & &0& & &}   
~.}
If we were to compactify the theory, obtaining
$T^6/\Zop_2\times\Zop_2$ orbifolds, there would be more fixed points 
and correspondingly larger contributions from the twisted sectors. For
most of our analysis it will however be sufficient to consider the
non-compact situation.

\subsec{The theory without discrete torsion}

The theory without discrete torsion has  conventional
fractional Dirichlet branes
for which all $s_i$ are even. For the simplest case where $s_i=0$,
this brane is stuck at the fixed plane of the orbifold group, 
$x^3=\cdots=x^8=0$. All these branes are described by a superposition
of boundary states where we have a non-trivial component in every
closed string sector of the theory. These components are invariant
under the GSO- and the orbifold projection, and the branes are charged
with respect to the twisted and untwisted R-R potentials.

In addition to this conventional
fractional Dirichlet brane, the theory also has 
supersymmetric `projective fractional' D(r;1,1,1) branes. For a fixed
orientation, the moduli space of the brane consists of three branches,
namely the fixed planes of $g_1$, $g_2$ and $g_3=g_1g_2$. The boundary
state description of the brane is slightly different for the different
branches of the moduli space, and we shall in the following only give
the explicit formula for the case of the $g_1$ branch. (The relevant
modifications for the other branches are self-evident.) Let us denote 
by ${\bf y}$ the position of the brane in the directions that are
unaffected by the orbifold action, and by ${\bf a}$ the coordinates in
the $x^3,x^4$ directions on the fixed plane of $g_1$. The relevant
boundary state is then of the form 
\eqn\pf{\eqalign{|D(r;& s_1,s_2,s_3); {\bf y},{\bf a},\epsilon,
\epsilon'\rangle
= |D(r;s_1,s_2,s_3);{\bf y},{\bf a}\rangle_{{\rm NS-NS;U}} 
+\epsilon |D(r;s_1,s_2,s_3);{\bf y},{\bf a}\rangle_{{\rm R-R;U}} \cr
& +\epsilon'\left(
 |D(r;s_1,s_2,s_3);{\bf y},{\bf a}\rangle_{{\rm NS-NS;T}_{g_1}} 
+ \epsilon 
|D(r;s_1,s_2,s_3);{\bf y},{\bf a}\rangle_{{\rm R-R;T}_{g_1}}\right) 
\cr
& + |D(r;s_1,s_2,s_3);{\bf y},{\bf - a}\rangle_{{\rm NS-NS;U}}  
+\epsilon |D(r;s_1,s_2,s_3);{\bf y},{\bf - a}\rangle_{{\rm R-R;U}}
\cr 
& -\epsilon'\left(
 |D(r;s_1,s_2,s_3);{\bf y},{\bf - a}\rangle_{{\rm NS-NS;T}_{g_1}}  
+ \epsilon 
|D(r;s_1,s_2,s_3);{\bf y},{\bf -a}\rangle_{{\rm R-R;T}_{g_1}}\right)
\,,}}
where $s_1=s_2=s_3=1$ and $\epsilon,\epsilon'=\pm 1$. Here and in the 
following we always restrict
${\bf a}$ to parametrise the `reduced' space; in the present
case this means that ${\bf a}$ parametrises for example the half-space
characterised by $a_3\geq 0$. It was shown in \gab\ that this state is
invariant under the GSO- and orbifold projection, and that it gives 
rise to the projective representation of the orbifold group 
\eqn\repsorig{\eqalign{\gamma(g_1)& = 
\left( \matrix{1&0 \cr 0 & -1}\right) \cr
\gamma(g_2)& = 
\left( \matrix{0&\pm 1 \cr 1 & 0}\right) \cr
\gamma(g_3)& = 
\left( \matrix{0&\mp 1 \cr 1 & 0}\right) \cr
\gamma((-1)^F) & = \left( \matrix{1&0 \cr 0 & 1}\right)
}}
on the $2\times 2$ Chan-Paton indices of the open strings that begin
and end on the D(r;1,1,1) brane. 

While the boundary state contains components from the $g_1$-twisted
sector, it is not charged with respect to any massless R-R field in the
$g_1$-twisted sector; this is simply a consequence of the fact that
the massless ground state of the boundary state (with zero momentum)
is independent of ${\bf a}$, and that the above boundary state
consists of the difference between the state at ${\bf a}$ and the one
at $-{\bf a}$. However, the above boundary states are charged under
the massless R-R fields in the untwisted sector. In fact, there are
eight different such massless fields that correspond to the eight
different orientations of the D(r;1,1,1) brane along the internal
directions; all of these D-branes are consistent. 

One may wonder whether it is consistent to have both a conventional
fractional brane and a projective fractional brane in one theory. In  
particular, one may ask whether the action of the orbifold group on
the open string that begins on the conventional
fractional brane and ends on the
projective fractional brane respects the relations of the orbifold
group.\footnote{$^\ddagger$}{This issue was not analysed in \gab. We
thank Greg Moore for drawing our attention to this point.} As we shall
see momentarily, this is indeed consistent. Actually, one could
have turned the argument around and predicted that the D(r;1,1,1)
brane has to carry a projective representation of the orbifold group
from the consistency analysis of this open string. 

The open string between the conventional fractional brane and the
projective fractional brane has two Chan-Paton indices that label on
which of the two D(r;1,1,1) branes the open string begins (or
ends). Under the action of the orbifold group, this 2-vector
transforms in the projective representation of the orbifold group that
characterises the D(r;1,1,1) brane. In order for the whole action of
the orbifold group to be consistent we therefore have to have that the
action on the oscillator states of the corresponding open string is
also projective. For any pair of a conventional fractional and a projective
fractional brane, the open string has three fermionic zero modes in
the internal directions, and an odd number of fermionic zero modes in
the directions that are unaffected by the action of the orbifold
group. (This is true for both the NS and the R sector.) Of the three 
fermionic zero modes in the internal directions, one is always in the
$x^3-x^4$ plane, one in the $x^5-x^6$ plane and one in the $x^7-x^8$
plane; for definiteness, let us assume (without loss of generality)
that the relevant fermionic zero modes are $\psi^3_0$, $\psi^5_0$ and
$\psi^7_0$. The action of $g_1$ and $g_2$ on the ground states of the
open string is then given by
\eqn\action{\eqalign{ g_1 & = \pm 2 i \psi^5_0 \psi^7_0 \cr 
                      g_2 & = \pm 2 i \psi^3_0 \psi^7_0\,,}}
where the prefactor has been fixed (up to a sign) by the condition
that $g_1^2=g_2^2=1$. (We are assuming here that the fermionic zero
modes satisfy the Clifford algebra
$\{\psi^\mu_0,\psi^\nu_0\}=\delta^{\mu,\nu}$.) Irrespective of the
choices for the signs, we then have the identity
\eqn\comm{ g_1 g_2 = - g_2 g_1 \,,}
and this implies that the action of the orbifold group on the
oscillator states of the open string is also projective. Taken
together with the projective representation on the Chan-Paton indices,
the whole action is then a conventional representation, as has to be
the case for consistency. Thus we have seen that it is necessary for 
consistency that the D(r;1,1,1) has a projective representation of the
orbifold group.

It may be worthwhile to point out that the above D-brane spectrum
falls slightly outside the framework described in \refs{\gimpol}. As
we mentioned in the introduction, the action of the orbifold group on
the open string space of states can be written as 
\eqn\total{ g |\psi,ij\rangle  
= \gamma(g)_{ii'} |U(g)\psi,i'j'\rangle\, \gamma(g)^{-1}_{j'j} \,,}
where $\psi$ is an element of the open string Hilbert space,
$\psi\in{\cal H}_{ij}$ (that depends in general on the Chan-Paton
indices $ij$), the action on the Chan-Paton indices is given by the
representation $\gamma$ while that on the open string states in 
${\cal H}=\oplus_{ij} {\cal H}_{ij}$ is described by $U$. If we assume
(as was done in \refs{\gimpol}) that $U$ defines a conventional
representation of the orbifold group, it then follows (using the
factorisation properties of the open string diagrams) that $\gamma$
must define a conventional or a projective representation of the
orbifold group. What we have found above is that $\gamma$ is a direct
sum of projective and conventional representations of the orbifold
group (and therefore {\it neither} projective nor conventional). This
is consistent with the arguments of \refs{\gimpol} because $U$ does
not define a conventional representation of the orbifold group on the
whole space of open string  states. (Indeed, as we have just shown,
$U$ acts for example projectively in the sector of the open strings
between D(r;0,0,0) and D(r';1,1,1), while it describes a conventional
representation in all sectors describing strings that begin and end on
the same brane.) In essence, the consistency analysis of
\refs{\gimpol} amounts to checking that the total orbifold action
defines a conventional representation for every open string sector;
this is what we have verified above, and what we shall analyse in the
following. 
\bigskip

It is not difficult to see that the above branes are the only
supersymmetric branes of the theory and that they account already for
all R-R charges of the theory. In addition to these BPS branes, the
theory also has a number of non-BPS branes. One of them is yet another
kind of projective fractional brane for which one of the three $s_i$
is even, while the other two are odd. This brane has been discussed
before in \stef, but it was not realised there that it gives rise 
to a projective representation of the orbifold group (as we shall see 
momentarily). For definiteness, let us assume that $s_1$ is even;
the boundary state of this brane is then given by 
\eqn\pfp{\eqalign{|D(r;& s_1,s_2,s_3); {\bf y},{\bf a},\epsilon,
\epsilon'\rangle
= |D(r;s_1,s_2,s_3);{\bf y},{\bf a}\rangle_{{\rm NS-NS;U}} 
+\epsilon |D(r;s_1,s_2,s_3);{\bf y},{\bf a}\rangle_{{\rm R-R;U}}  \cr
& + \epsilon'\left( 
|D(r;s_1,s_2,s_3);{\bf y},{\bf a}\rangle_{{\rm NS-NS;T}_{g_1}} 
+ \epsilon 
|D(r;s_1,s_2,s_3);{\bf y},{\bf a}\rangle_{{\rm R-R;T}_{g_1}}\right)
\cr
& +|D(r;s_1,s_2,s_3);{\bf y},{\bf - a}\rangle_{{\rm NS-NS;U}}
-\epsilon|D(r;s_1,s_2,s_3);{\bf y},{\bf - a}\rangle_{{\rm R-R;U}}
\cr 
&- \epsilon'\left(
|D(r;s_1,s_2,s_3);{\bf y},{\bf - a}\rangle_{{\rm NS-NS;T}_{g_1}} 
- \epsilon
|D(r;s_1,s_2,s_3);{\bf y},{\bf -a}\rangle_{{\rm R-R;T}_{g_1}}\right)
\,,}}
where now $s_1$ is even and $s_2=s_3=1$. It is easy to see from
table~2 of \gab\ that this state is invariant under the action of the
GSO projection and the orbifold group. The corresponding D-brane is
charged under a massless R-R field from the $g_1$-twisted sector. It
is also clear that this boundary state reduces to the expression given
in \stef\ as ${\bf a}\rightarrow 0$. 

The determination of the corresponding projection operators in the
open string requires a little bit of care. First of all, since the two
copies of the brane (at ${\bf a}$ and $-{\bf a}$) have opposite bulk R-R 
charge, the
action of $(-1)^F$ in the open string involves a non-trivial action on
the Chan-Paton factors which is given by (conjugation with) 
\eqn\sigma{ \gamma\Bigl((-1)^F\Bigr) 
= \left( \matrix{1&0 \cr 0 & -1}\right)\,.}
In addition, the action of the orbifold group on the Chan-Paton
indices is given by the matrices 
\eqn\reps{\eqalign{\gamma(g_1)& = 
\left( \matrix{1&0 \cr 0 & -1}\right) \cr
\gamma(g_2)& = 
\left( \matrix{0&\pm 1 \cr 1 & 0}\right) \cr
\gamma(g_3)& = 
\left( \matrix{0&\mp 1 \cr 1 & 0}\right)\,,}}
as can be read off from the boundary state (this is described in some
detail in \gab). As before, this is a projective representation of the
orbifold group, but now also the action of $(-1)^F$ does not commute
any more with the action of $g_2$ and $g_3$: in fact we have 
\eqn\novel{\eqalign{
\gamma(g_1) \gamma\Bigl((-1)^F\Bigr) & = 
\gamma\Bigl((-1)^F\Bigr) \gamma(g_1) \cr
\gamma(g_2) \gamma\Bigl((-1)^F\Bigr) & = 
-\gamma\Bigl((-1)^F\Bigr) \gamma(g_2) \cr
\gamma(g_3) \gamma\Bigl((-1)^F\Bigr) & = 
-\gamma\Bigl((-1)^F\Bigr) \gamma(g_3)\,.}}

As before, we need to check whether these non-BPS branes are mutually
consistent with the BPS branes that we have described above, \ie\
whether the total action on the open strings between a BPS brane and a
non-BPS brane respects the relations of the orbifold group and of
$(-1)^F$. It is not difficult to see that this is indeed the
case. Consider, for instance, the NS sector of the open string between
a projective fractional D(r;0,1,1) brane extended along the $x^5$ and
$x^7$ directions and a conventional
fractional D(r';0,0,0) brane. The fermion zero
modes in the orbifold directions are $\psi_0^5$ and $\psi_0^7$, and
the action of the group elements on the ground states of the open
string is then given by   
\eqn\actiontwo{\eqalign{ g_1 & = \pm 2i\psi_0^5\psi_0^7 \cr 
                         g_2 & = \pm\sqrt{2}\psi_0^7 \Gamma \cr
                         g_3 & = \pm\sqrt{2}\psi_0^5 \Gamma\,,}}
where $\Gamma$ is the chirality operator that is proportional to the
product of all fermionic zero modes. The total number of fermionic
zero modes is even and therefore $\Gamma$ commutes with
$(-1)^F$. This implies that both $g_2$ and $g_3$ anti-commute
with $(-1)^F$, thus providing precisely the signs that cancel the
signs in \novel. We also have that the orbifold operators among
themselves satisfy the relations of the projective representation of
the orbifold group; if we combine this action with the projective
action on the Chan-Paton factors \reps\ we get a conventional
representation of the orbifold group, as required by consistency. The
analysis for the R sector is analogous.  

The situation for the string between the non-BPS brane and the
projective fractional  D(r';1,1,1) brane is similar. Again, if we
consider the NS sector of the open string between a projective
fractional D(r;0,1,1) brane  along the $x^5$ and $x^7$ directions and
a projective fractional D(r';1,1,1) brane along the $x^3$, $x^5$ and
$x^7$ directions, we have only one fermionic zero mode, namely
$\psi_0^3$. The group elements are then represented on the ground
states of the open string by  
\eqn\actionthree{\eqalign{ g_1 & = 1 \cr 
                           g_2 & = \pm\sqrt{2}\psi_0^3 \Gamma\cr
                           g_3 & = \pm\sqrt{2}\psi_0^3 \Gamma\,,}}
where $\Gamma$ is again the chirality operator.
Again both $g_2$ and $g_3$ anti-commute with $(-1)^F$, but now the
orbifold generators commute among themselves and therefore define a
conventional representation of the orbifold group. The Chan-Paton
indices of both branes transform in the same (projective)
representation of the orbifold group, and thus the action on the
Chan-Paton indices is a conventional representation of the orbifold
group. Together with the above (conventionial) representation on the
oscillator states, the total action on the open string states
therefore satisfies the relations of the orbifold group. The same also
applies for the commutation relations between $(-1)^F$ and the
orbifold generators. 

There is yet another consistency condition that we should analyse, but
that is somewhat more subtle. In the above, we have only  analysed  
the D(r;0,1,1) brane (or the D(r;2,1,1) brane), but we have
implicitly claimed that the theory also has branes where $s_2$ or
$s_3$ is even while the other two $s_j$ are odd. In each case, we can
repeat the above analysis and thereby demonstrate that each of these
branes gives rise to a consistent string between the non-BPS brane in 
question and any BPS brane. However, we also need to check whether the
open string between two different such non-BPS branes is
consistent. As an example, let us consider the open string between
the D(r;0,1,1) brane and the D(r;1,0,1) brane. The action of the
orbifold group on the Chan-Paton indices of the two branes is the
same, giving rise to a conventional representation of the orbifold
group on the Chan-Paton factors; in order for the whole open string to
be consistent we therefore have to have that the action on the
oscillator states is also a conventional representation of the
orbifold group. 

Unfortunately, the precise action of the orbifold group on the open
string oscillator states is not straightforward to determine, and we
are therefore not able to check this consistency condition directly. 
The difficulty in establishing the orbifold action on the open string
states is due to the fact that the moduli spaces of the two branes in
question are different: the moduli space of the D(r;0,1,1) brane
is the fixed plane of $g_1$, while that of the D(r;1,0,1) is the fixed
plane of $g_2$. Under the action of $g_1$ the open string that begins
at the D(r;0,1,1) localised at ${\bf a}=(a_3,a_4)$ and ends at the
D(r;1,0,1) localised at ${\bf b}=(b_5,b_6)$ is mapped to a string that
begins at $(a_3,a_4)$ and ends at $(-b_5,-b_6)$. Since these two
strings are not parallel to one another, there is no canonical way in
which we can identify the corresponding Hilbert spaces, and therefore
no sense in which $g_1$ acts as a product of fermionic zero modes. The
situation is similar for the action of $g_2$. 

It may also be worthwhile to point out that if it {\it was} possible
to define the action of $g_i$ in terms of fermionic zero modes, our 
analysis would imply that the D(r;0,1,1) and the D(r;1,0,1) are
mutually inconsistent (which would seem somewhat unlikely). Indeed,
the R-sector of the open string between these two branes has (for a
suitable orientation of the branes) the zero modes $\psi^4_0$ and
$\psi^6_0$. If we could define $g_1=\sqrt{2}\psi^6_0\Gamma$ and
$g_2=\sqrt{2}\psi^4_0\Gamma$, we would have $g_1g_2=-g_2g_1$, and the total
orbifold action on the open string states would be inconsistent. 
\smallskip

In the uncompactified theory, the non-BPS D(r;0,1,1) brane is
unstable, but it becomes stable if we compactify the orbifolded
directions along which the brane wraps ($x^5$ and $x^7$, say) on
sufficiently small circles \stef. Indeed, it is clear from the
boundary state  \pfp\ that the string between the brane at ${\bf a}$
and the brane at $-{\bf a}$ (these strings correspond to the
off-diagonal Chan-Paton indices) has the `wrong' GSO-projection, and 
therefore that the tachyonic ground state survives the
GSO-projection. However, it is also clear from \reps\ that the
orbifold projection acts as $(1-g_1)/2$ on these open strings, and
therefore that the ground state tachyon is not orbifold invariant. On
the other hand, the open string state with momentum $p^5$ is mapped to
the state with momentum $-p^5$ under the action of $g_1$, and
therefore the anti-symmetric combination of these two states is
invariant under $g_1$. Furthermore, by considering a suitable linear
combination of the two off-diagonal Chan-Paton indices, the state can
be made invariant under the action of $g_2$, and therefore under the
action of the whole orbifold group. The resulting physical state is
tachyonic provided that $R^5$ is sufficiently large and the separation
between both copies of the brane is sufficiently small --- for
instance, for ${\bf a}=0$ the precise condition on $R^5$ is
$R^5>\sqrt{2 \alpha'}$; the non-BPS D-brane can therefore only be 
stable if $R^5$ is sufficiently small (\ie\ $R^5<\sqrt{2 \alpha'}$). 
In the regime where the D(r;0,1,1) brane is unstable it decays
presumably into two non-BPS D(r;0,0,1) branes of the type considered
in \stef,\footnote{$^\star$}{It may be worth pointing out that for 
the D(r;0,0,1) branes of \stef\ $r+1$ is odd in IIA and even in IIB;
thus these branes occur for the {\it same} value of $r$ as the non-BPS 
D(r;0,1,1) brane we have just discussed.} or into four fractional
D(r;0,0,0) branes, depending on $R^7$. Other instabilities arise if
the 6 and 8 directions are compactified on sufficiently small circles
because we obtain then tachyonic winding states.  

The non-BPS D(r';0,0,1) brane discussed in \stef\ is quite special to
$N=2$, and does not generalise to $N>2$. However, the theory has yet
another non-BPS D(r;0,0,1) brane (where now $r+1$ is even in IIA and
odd in IIB) which will generalise to $N>2$. If we require that the
open strings between this brane and the BPS branes are consistent, an
analogous analysis to the above implies that the action of the
orbifold group on the Chan-Paton indices must be a conventional
representation, but that $(-1)^F$ must act non-trivially so that it
anticommutes with $g_1$ and $g_2$ (but not $g_3$). A boundary state
that gives rise to this action is of the form    
\eqn\pfpp{\eqalign{
|D(r;& s_1,s_2,s_3); {\bf y},{\bf c},\epsilon,\epsilon'\rangle
= |D(r;s_1,s_2,s_3);{\bf y},{\bf c}\rangle_{{\rm NS-NS;U}} 
+\epsilon |D(r;s_1,s_2,s_3);{\bf y},{\bf c}\rangle_{{\rm R-R;U}}  \cr
& + \epsilon'\left( 
|D(r;s_1,s_2,s_3);{\bf y},{\bf c}\rangle_{{\rm NS-NS;T}_{g_3}} 
+ \epsilon 
|D(r;s_1,s_2,s_3);{\bf y},{\bf c}\rangle_{{\rm R-R;T}_{g_3}}\right)
\cr
& +|D(r;s_1,s_2,s_3);{\bf y},{\bf - c}\rangle_{{\rm NS-NS;U}}
-\epsilon|D(r;s_1,s_2,s_3);{\bf y},{\bf - c}\rangle_{{\rm R-R;U}}
\cr 
&+ \epsilon'\left(
|D(r;s_1,s_2,s_3);{\bf y},{\bf - c}\rangle_{{\rm NS-NS;T}_{g_3}} 
- \epsilon
|D(r;s_1,s_2,s_3);{\bf y},{\bf -c}\rangle_{{\rm R-R;T}_{g_3}}\right)
\,,}}
where $s_1=s_2=0$, $s_3=1$ and ${\bf c}$ parametrises now the fixed
plane of $g_3$. It is again easy to see that this boundary state is
invariant under the GSO- and the orbifold projection. However, because
of the relative minus signs, the resulting D-brane is uncharged with
respect to any R-R potential. 

As before, we can read off from the above boundary state the action of 
the various operators on the Chan-Paton indices
\eqn\novelone{\eqalign{\gamma(g_1)& = 
\left( \matrix{0&1 \cr 1 & 0}\right) \cr
\gamma(g_2)& = 
\left( \matrix{0&1 \cr 1 & 0}\right) \cr
\gamma(g_3)& = 
\left( \matrix{1&0 \cr 0 & 1}\right) \cr
\gamma((-1)^F) & = \left( \matrix{1&0 \cr 0 & -1}\right)\,.}}
This defines indeed a conventional representation of the orbifold
group for which $(-1)^F$ anti-commutes with $g_1$ and $g_2$.

Incidentally, a similar consistency analysis also applies to the brane
proposed in \refs{\stef}. The main difference to the situation above
is that the overall number of fermionic zero modes for the strings
between the non-BPS brane and the BPS branes is odd in that case. In
order to be able to define the chirality operator $\Gamma$ (that
enters in the definition of the orbifold operators as in \actiontwo\
and \actionthree) it is then necessary to introduce an additional
`boundary' fermion, as discussed in a similar situation by Witten
\refs{\wittenK}. This procedure doubles the degrees of freedom of the
open string, and allows for an action of the matrices \novelone\ on
the (two-dimensional) space of multiplicities. The resulting open
string loop amplitudes are then in agreement with those that follow
from the boundary state given in \refs{\stef}.  

We also have to check that the non-BPS branes for which one $s_i$ is
odd (while the other two $s_j$ are even) are consistent with
themselves and the other non-BPS branes. In those cases where the
relevant branes are defined on the same branch, the orbifold
generators can be expressed in terms of fermionic zero modes, and we
have verified, using similar arguments as above, that the branes are
indeed consistent. In the other cases, the situation is more
complicated, and we do not know how to check this consistency
condition directly. 

Finally, it also follows from \novelone\ that the above non-BPS
D(r;0,0,1) brane is unstable, irrespective of whether we compactify or 
not. Indeed, the off-diagonal components of the Chan-Paton matrix
correspond to strings with the wrong GSO-projection, and therefore
contain a tachyonic ground state in the NS sector. Since $g_3$ acts as
the identity matrix, both these states are invariant under $g_3$; a
certain linear combination of them is thus invariant under $g_1$ and
$g_2$, and therefore defines a physical tachyonic state in the open
string.   

For the convenience of the reader, let us summarise the D-branes whose
boundary states we have discussed above.
\vskip 0.4cm
\vbox{
\centerline{\vbox{
\hbox{\vbox{\offinterlineskip
\def\tablespace{height2pt&\omit&&\omit&&\omit&\cr}

\def\tableruleA{\tablespace\noalign{\hrule height1pt}\tablespace}
\hrule\halign{&\vrule#&\strut\hskip0.2cm\hfil#\hfill\hskip0.2cm\cr
\tablespace
& && BPS/non-BPS && type of representation  &\cr
\tableruleA
& D(r;even,even,even) && BPS && conventional &\cr \tablespace
& D(r;1,1,1) && BPS && projective \repsorig\  &\cr \tablespace
& D(r;even,1,1) && non-BPS && projective \sigma\ and \reps\ &\cr
\tablespace 
& D(r;even,even,1) && non-BPS && `conventional' \novelone\
&\cr \tablespace
\tablespace}\hrule}}}}
\centerline{
\hbox{{\bf Table 1:} {\it The D-brane spectrum of the 
$\Zop_2\times \Zop_2$ orbifold without discrete torsion.}}}}
\vskip 0.5cm
In the above it is understood that all permutations of `even' and $1$
are included, and that  $r$ is determined in terms of $s_i$ as
discussed before. In the last case, the action of the orbifold group
is a conventional representation, but it does not commute with
$(-1)^F$.

\subsec{The theory with discrete torsion}

For the theory with discrete torsion, the r\^oles of the conventional
and projective fractional branes are reversed. For $s_i$ even, the
theory now has a projective fractional brane. Again, its moduli space
consists of the three fixed planes of $g_1$, $g_2$ and
$g_3=g_1g_2$. In the $g_1$-branch the relevant boundary state is then
given by \pf\ where 
now $s_i$ is even. As before, this state is invariant under the GSO-
and orbifold projection, and it gives rise to a projective
representation of the orbifold group on the $2\times 2$ Chan-Paton
indices of the open strings that begin and end on this brane. 

In the limit ${\bf a}\rightarrow 0$, this boundary state reduces
to what has been described before in \diagom\ (see also
\bcf), but the above description is more general, and in particular
describes the relevant boundary state for all points on its moduli
space. The fact that the boundary state involves components from the 
$g_1$-twisted sector suggests that the brane cannot move off
the fixed planes that describe its moduli space. 

The theory also has a conventional fractional D(r;1,1,1) brane. The
corresponding boundary state has components in all closed string
sectors of the theory; it is given by
\eqn\pfNtwoDT{\eqalign{|D(r;1,1,1);&\, {\bf y},\epsilon,
\epsilon_1,\epsilon_2\rangle=\cr
&|D(r;1,1,1);{\bf y}\rangle_{{\rm NS-NS;U}} 
+\epsilon |D(r;1,1,1);{\bf y}\rangle_{{\rm R-R;U}}  \cr
& + \epsilon_1 \Bigl(
 |D(r;1,1,1);{\bf y}\rangle_{{\rm NS-NS;T}_{g_1}}
+ 
\epsilon |D(r;1,1,1);{\bf y}\rangle_{{\rm R-R;T}_{g_1}} \Bigr)
\cr
& + \epsilon_2 \Bigl(
 |D(r;1,1,1);{\bf y}\rangle_{{\rm NS-NS;T}_{g_2}}
+ 
\epsilon |D(r;1,1,1);{\bf y}\rangle_{{\rm R-R;T}_{g_2}} \Bigr)
\cr
& + \epsilon_1 \epsilon_2 \Bigl(
 |D(r;1,1,1);{\bf y}\rangle_{{\rm NS-NS;T}_{g_3}}
+ 
\epsilon |D(r;1,1,1);{\bf y}\rangle_{{\rm R-R;T}_{g_3}} \Bigr)
\,.}} 
Because of the same argument as above, the open
string between the two types of branes is then again consistent. Also,
these are the only supersymmetric branes, and they account for all R-R
charges of the theory. 

There is now a consistent D-brane for which one of the $s_i$ is odd,
while the other two are even. If $s_1=1$, the relevant boundary state
is described by \pfp. This brane carries twisted R-R charge and is
stable (for a certain regime of radii in the compactified theory) but 
non-BPS. Finally, the theory has an unstable, uncharged, non-BPS
D-brane for which precisely one $s_i$ is even. The consistency and
stability analysis is as in the case without discrete torsion. All of
this is in agreement with T-duality that relates the theory with and
without discrete torsion \refs{\vafawitten}. The D-brane spectrum can
be summarised by 
\vskip 0.4cm
\vbox{
\centerline{\vbox{
\hbox{\vbox{\offinterlineskip
\def\tablespace{height2pt&\omit&&\omit&&\omit&\cr}

\def\tableruleA{\tablespace\noalign{\hrule height1pt}\tablespace}
\hrule\halign{&\vrule#&\strut\hskip0.2cm\hfil#\hfill\hskip0.2cm\cr
\tablespace
& && BPS/non-BPS && type of representation  &\cr
\tableruleA
& D(r;even,even,even) && BPS && projective  \repsorig\ &\cr
\tablespace 
& D(r;1,1,1) && BPS && conventional  &\cr \tablespace
& D(r;even,1,1) && non-BPS && `conventional' \novelone\ &\cr
\tablespace 
& D(r;even,even,1) && non-BPS && projective \sigma\ and \reps\
&\cr \tablespace
\tablespace}\hrule}}}}
\centerline{
\hbox{{\bf Table 2:} {\it The D-brane spectrum of the 
$\Zop_2\times \Zop_2$ orbifold with discrete torsion.}}}}
\vskip 0.7cm

The main aim of this paper is to explain how the above analysis
generalises to the case of a general $\Zop_N\times\Zop_N$ orbifold. In
doing so, we shall encounter a number of interesting and new
phenomena. As shall become apparent, the analysis depends on
whether $N$ is odd or even, and further on whether an even $N$ is
divisible by four or not. After mentioning some generalities, we shall
discuss these  different cases separately in the following.

\newsec{The $\Zop_N\times\Zop_N$ orbifold: generalities}

Let us consider the $\Zop_N\times\Zop_N$ orbifold of $\Rop^6$. We 
identify $\Rop^6\simeq \Cop^3$ by defining 
\eqn\define{\eqalign{
z_1 & = x^3 + i x^4 \cr
z_2 & = x^5 + i x^6 \cr
z_3 & = x^7 + i x^8\,.}}
The two cyclic groups then act on $z_i$ by 
\eqn\action{\eqalign{
g_1 (z_1,z_2,z_3) & = (z_1, e^{-{2\pi i \over N}} z_2, 
                          e^{{2\pi i\over N}} z_3) \cr
g_2 (z_1,z_2,z_3) & = (e^{{2\pi i\over N}} z_1, z_2, 
                          e^{-{2\pi i\over N}} z_3)\,.}}

The possible discrete torsion theories are classified by
$H^2(\Zop_N\times\Zop_N;U(1))= \Zop_N$. Indeed, the possible
discrete torsion phases $\epsilon(g,h)$ are determined in terms of 
\eqn\dis{\omega\equiv \epsilon(g_1,g_2) = e^{{2\pi i m\over N}} \qquad
\hbox{where $m=0,\dots,N-1$.}}
This fixes all the phases $\epsilon(g,h)$ since we have the relations 
\eqn\cons{\eqalign{ 
\epsilon(h_1 h_2,k) & = \epsilon(h_1,k) \epsilon(h_2,k)  \cr
\epsilon(h,g) & = \epsilon(g,h)^{-1}\,. }}
In this paper, we shall only consider `minimal' discrete torsion,
which means that  $\omega$ is a generator of $\Zop_N$.

Let us also give the Hodge diamond that  summarises the spectrum of
R-R ground states. In the theory without discrete torsion, the
untwisted sector contributes (for $N>2$) \refs{\dougfiol}
\eqn\hodgeN{
\matrix {& & & &1& & &  \cr
           & & &0& &0& &  \cr
           & &0& &3& &0&  \cr
           &1& &0& &0& &1 \cr
           & &0& &3& &0&  \cr
           & & &0& &0& &  \cr
           & & & &1& & &  } 
}
to the Hodge diamond. The contribution of the twisted sectors is
\eqn\hodgetwN{
\matrix {& & & &0& & &  \cr
           & & &0& &0& &  \cr
           & &0& &{(N+4)(N-1)\over 2}& &0&  \cr
           &0& &0& &0& &0 \cr
           & &0& &{(N+4)(N-1)\over 2}& &0&  \cr
           & & &0& &0& &  \cr
           & & & &0& & &   }
~.}
In the theory with minimal discrete torsion, the untwisted
contribution is the same, while the twisted sectors now contribute
\eqn\hodgetwdt{
\matrix {& & & &0& & &  \cr
           & & &0& &0& &  \cr
           & &0& &0& &0&  \cr
           &0& &3& &3& &0 \cr
           & &0& &0& &0&  \cr
           & & &0& &0& &  \cr
           & & & &0& & &   }
~.}
The BPS D-branes that have been constructed in the literature
\refs{\dougfiol,\diagom} do not couple to these twisted R-R
potentials; it is one of the aims of this paper to construct 
D-branes that carry these charges.

\newsec{The $\Zop_N\times\Zop_N$ orbifold for odd $N$}
\subsec{The theory without discrete torsion}

First of all, as in the $N=2$ case studied in Section 2, the theory
has conventional fractional BPS D-branes for which all $s_i$ are
even. These branes are charged under twisted and untwisted R-R
potentials. The corresponding boundary states are given by 
\eqn\fb{\eqalign{&
|D(r;s_1,s_2,s_3); {\bf y},\epsilon,\epsilon_1,\hat{\epsilon}_1\rangle
= \cr & \sum_{m,n=0}^{N-1} \epsilon_1^m \hat{\epsilon}_1^n
\left(|D(r;s_1,s_2,s_3);{\bf y},{\bf 0}\rangle_{
     {\rm NS-NS;T}_{g_1^mg_2^n}}  
+\epsilon |D(r;s_1,s_2,s_3);{\bf y},{\bf 0}\rangle_{
     {\rm R-R;T}_{g_1^mg_2^n}}
\right)\,,}}
where, for notational convenience, we have considered the untwisted
sector to be the sector twisted by the trivial group element.

For the remaining branes, there are substantial differences compared
to the $N=2$ case. In particular, when some of the $s_i$ are odd, the
orbifold group maps a copy of the brane in the covering space to other
copies with different orientations. For instance, to build an
invariant configuration of D(r;1,1,1) branes in the covering space,
we now need at least $N^2$ copies, which can therefore support a
regular representation of the orbifold group which usually corresponds
to a bulk brane. This brane carries untwisted R-R charge and is in
fact BPS (using the techniques of \refs{\bdl} one can check that the
branes-at-angles configuration on the covering space is such that it
preserves 1/4 supersymmetry), but does not carry any twisted R-R
charge.   

As we have explained before, the total action of the orbifold group on
the open string space of states is given by \total. Consistency
requires that this defines a conventional representation of the
orbifold group. A general multi-point amplitude of some open string
states can be factorised into the amplitude of the open string vertex
operators, and a trace over the Chan-Paton indices. It is then
conventional to define $U(g)$ on the different open string states so
that the amplitude of the open string vertex operators is invariant
under the group action. Since the total action has to be a
conventional representation of the orbifold group, this fixes then the 
action of the orbifold group on the Chan-Paton indices.

Sometimes the action of $U(g)$ on the open string Hilbert spaces is
canonically defined (such as in the case $N=2$ we discussed in
Section~2), but in general this may not be the case, in particular, if
the group elements do not map the open strings to parallel
strings. In such cases, the invariance of the amplitudes of the
vertex operators fixes in principle the action of $U(g)$, but it is
difficult to determine the precise action in practice.  

One example for which this discussion is relevant is the bulk
D(r;1,1,1) brane we have just considered. In this case, the different
open strings are not mapped into parallel strings under the action of
the orbifold group, and there is therefore no canonical way to define
the group elements on the open string space of states. It is therefore
difficult to decide whether the action of the orbifold group on the
Chan-Paton factors is a conventional or a projective representation in
this case. 
\medskip

As we have argued above, the D(r;1,1,1) brane is a bulk brane rather 
than a fractional brane. Another way to see this is to observe that it
is impossible to write down a D(r;1,1,1) boundary state in any twisted  
sector:\footnote{$^\dagger$}{We thank Fred Roose for a discussion on
this point.} since $N$ is odd, the oscillators in each twisted sector 
are not half-integer moded for at least two of the three complex
directions $x^{2j+1}+ix^{2j+2}$. If this is the case, the only
boundary condition that has a non-trivial solution is DD or NN (where
the two letters refer to the boundary conditions for $x^{2j+1}$ and
$x^{2j+2}$, respectively.) On the other hand, the D(r;1,1,1) brane has
a mixed DN boundary condition for each of the three complex
directions.  

However, this argument does not exclude the existence of other
fractional D-branes. In fact, the theory has a conventional fractional
(non-BPS) D-brane for which exactly one $s_i$ is odd. For instance, a
brane with $s_1=1$ can have components in the sectors twisted by
$g_1^m$, because these group elements do not shift the modings in the
complex direction for which we have a mixed (DN) boundary condition
($x^3+ix^4$). The boundary state corresponding to such a brane is
\eqn\bsonezerozero{|D(r;s_1,s_2,s_3);{\bf y},{\bf a},\epsilon,\epsilon_1
\rangle= 
\sum_{m=0}^{N-1} \epsilon_1^m \sum_{n=0}^{N-1} 
|D(r;s_1,s_2,s_3);{\bf y},g_2^n {\bf a},
\epsilon\rangle_{{\rm T}_{g_1^m}}\,,} 
where it is understood that the orientations of the different
component branes are such that the total configuration is orbifold
invariant. Here we have used the short-hand notation
\eqn\short{\eqalign{
|D(r;s_1,s_2,s_3);& {\bf y},{\bf a},\epsilon\rangle_{{\rm T}_{g_1^m}}
\cr 
& = |D(r;s_1,s_2,s_3);{\bf y},{\bf a}\rangle_{{\rm NS-NS;T}_{g_1^m}}
+ \epsilon 
|D(r;s_1,s_2,s_3);{\bf y},{\bf a}\rangle_{{\rm R-R;T}_{g_1^m}}  \,.}}
For consistency with the group relations, $\epsilon_1$ has to satisfy
$\epsilon_1^N=1$. The associated $N\times N$ matrices defining the
action on Chan-Paton factors are then 
\eqn\gammarep{
\gamma(g_1)={\rm diag}(\epsilon_1,\epsilon_1,\ldots,\epsilon_1)\,;\ \ 
\gamma(g_2)=\pmatrix{0&0&\cdots&0&1
\cr
1&0&\cdots&0&0
\cr
0&1&\cdots&0&0
\cr
\vdots&\vdots&&\vdots&\vdots
\cr
0&0&\cdots&1&0}\,.}
This defines indeed a conventional representation of the orbifold
group. The D-brane is non-supersymmetric --- the branes-at-angles
configuration in the covering space does not satisfy the criteria to
preserve any supersymmetry, see for instance \refs{\bdl,\polvtwo} ---
and, in fact, unstable (as in the $N=2$ case, one can show that the
open string spectrum contains tachyonic modes). These branes do not
carry any R-R charges: although the boundary state has R-R components
in the untwisted and the $g_1^m$-twisted sectors, there is no coupling
to a massless R-R potential (it is projected out by the subgroup
generated by $g_2$).  

As before, we should check whether the various open strings carry
consistent representations of the orbifold group and $(-1)^F$. 
Unfortunately, this is again very difficult to do explicitly since
none of the orbifold generators maps any of the constituent branes to
a parallel brane, and therefore, none of them can be represented by
fermionic zero modes.  

Apart from the bulk D(r;1,1,1) branes for which the representation of
the orbifold group is not easily determined, the branes that we
have considered above all transform in a conventional representation
of the orbifold group. This is to be contrasted with the situation for 
$N=2$ where the theory without discrete torsion also has branes that 
transform in a projective representation of the orbifold group. On
the other hand, for those branes for which we can unambiguously
identify the action of the orbifold group on the Chan-Paton indices
(namely the D(r;0,0,0) brane and the D(r';1,0,0) brane) the results
for $N=2$ and odd $N>2$ agree.

\subsec{The theory with minimal discrete torsion}

Like the $N=2$ theory with discrete torsion, the theory has
projective fractional D-branes where all $s_i$ are even. These branes
carry untwisted R-R charge and are BPS. The moduli space consists of
three different branes, and the boundary state for the $g_1$ branch
is given by  
\eqn\pbsodd{
|D(r;s_1,s_2,s_3);{\bf y},{\bf a},\epsilon,\epsilon_1\rangle=
\sum_{m=0}^{N-1}\epsilon_1^m \sum_{n=0}^{N-1} \omega^{-mn} 
|D(r;s_1,s_2,s_3);{\bf y},g_2^n{\bf a},
      \epsilon\rangle_{{\rm T}_{g_1^m}}\,, 
}
where we have used \short\ again. The powers of the discrete torsion
phase $\omega$ are determined by the condition that the boundary state
must be invariant under the action of the orbifold
group.\footnote{$^\ddagger$}{We are using here the convention that in
the theory with discrete torsion, the action of $g_i$ on the sector
twisted by $g_j$ is modified by multiplication with
$\epsilon(g_i,g_j)$; see \refs{\gab}.} The associated $N\times N$
matrices that define the action of the orbifold group on the
Chan-Paton indices are now given as  
\eqn\gammareppro{
\gamma(g_1)={\rm diag}(\epsilon_1,\epsilon_1 \omega^{-1},\ldots,
\epsilon_1 \omega^{-(N-1)})\,;\ \ 
\gamma(g_2)=\pmatrix{0&0&\cdots&0&1
\cr
1&0&\cdots&0&0
\cr
0&1&\cdots&0&0
\cr
\vdots&\vdots&&\vdots&\vdots
\cr
0&0&\cdots&1&0}\,.}
They define a projective representation of the orbifold group that is
characterised by 
\eqn\projj{ \gamma(g_1) \gamma(g_2) 
      = \omega^{-1} \gamma(g_2) \gamma(g_1) \,.}
This is in agreement with what was found in \refs{\dougfiol,\gomis}.

In addition to these projective fractional branes, the theory has also
bulk D(r;1,1,1) branes, just as in the case without discrete
torsion. The bulk branes also carry untwisted R-R charge and are
BPS. As before, these branes cannot be fractional, and the corresponding
representation on the Chan-Paton indices is not easily determined. 

The above branes account for all of the untwisted R-R charges of the
theory. However, the theory with discrete torsion also has twisted R-R
charges \refs{\vafawitten,\dougfiol}. The branes that are charged with
respect to these are non-BPS D-branes for which precisely one of the
$s_i$ is equal to $1$ while the other two $s_j$ are even. For the case
$s_1=1$, the relevant boundary state takes the form 
\eqn\projbsonezerozero{|D(r;s_1,s_2,s_3);{\bf y},{\bf a},
\epsilon,\epsilon_1 \rangle=
\sum_{m=0}^{N-1} \epsilon_1^m \sum_{n=0}^{N-1} \omega^{-mn}   
|D(r;s_1,s_2,s_3);{\bf y}, g_2^n {\bf a},
\epsilon\rangle_{{\rm T}_{g_1^m}}\,.}
This leads to an open string with an $N\times N$ Chan-Paton matrix,
for which the action of $g_1$ and $g_2$ is given again by
\gammareppro. 

This projective fractional non-BPS D-brane is quite an unusual
object. It carries twisted R-R charge (unlike the situation in the 
theory without discrete torsion, the orbifold projection does not
remove the relevant massless R-R potential to which it couples) which
is {\it not} carried by any of the BPS branes of the
theory that we have constructed above.%
\footnote{$^*$}{Strictly speaking, we have not proven that these are
the only BPS branes in the theory; on the basis of our analysis we can
therefore not exclude the possibility that there are unknown BPS
branes that carry this twisted charge. On the other hand, the relevant
twisted R-R charge does not appear as a central charge in the
supersymmetry algebra, and there is therefore no intrinsic reason why
the BPS states should be charged under these potentials. Furthermore,  
examples of manifolds are known that have non-trivial two-cycles in
homology, but for which no two-cycle can be chosen to be
supersymmetric \refs{\pap}.}
 
Since none of the (known) BPS D-branes carry twisted R-R charge, one
may expect that the non-BPS D-brane (that does carry this charge) has
preferred stability properties, and this is indeed what we
find.\footnote{$^\dagger$}{The situation is in fact similar to what
was found in \refs{\gutperle}: the different non-BPS D-branes can
decay into one another but do not seem to decay into BPS brane
anti-brane pairs.} As an example let us analyse the stability of the
D(r;1,0,0) brane. Before the orbifold projection, the open string
between two different copies of the brane on the covering space has a
tachyonic ground state. However, it follows from \gammareppro\ that
this ground state is projected out by $g_1$ because $g_1$ multiplies
strings with off-diagonal Chan-Paton factors by a non-trivial
phase. One may think that an orbifold invariant tachyonic state could
be obtained by giving the open string momentum or winding. However,
since $g_1$ acts trivially on the 34 directions, this must necessarily
involve the 5678 directions. For the D(r;1,0,0) brane, the open string
has only winding modes in these directions, and they are infinitely
massive in the uncompactified theory. Thus the D(r;1,0,0) brane 
appears to be stable in the non-compact
orbifold.\footnote{$^\ddagger$}{The absence of a tachyonic mode does not 
imply in general that the D-brane is stable; for example, a D-brane
whose open string does not contain a tachyonic mode may be metastable 
\refs{\majsen}. In the present case, however, there is no reason to
suppose that the brane is only metastable.}
If we compactify, say, the 56 directions on a
sufficiently small torus,\footnote{$^\star$}{Strictly speaking, this 
only applies to $N=3$ since we cannot compactify the orbifold for any
other odd $N$.} we get a tachyonic winding state that presumably
signals the decay into a configuration of non-BPS D(r;1,2,0)
branes. Since the shape of the torus is fixed by the orbifold
symmetry, there are no intermediate configurations; this is in
agreement with the fact that none of the other branes that we have
constructed carry twisted R-R charge. 

In turn, a D(r;1,2,0) brane is unstable in the uncompactified theory
(because the open string has a tachyon with infinitesimal momentum in
the 56 directions), but becomes stable if the 56 directions are
compactified on a sufficiently small torus.

\newsec{The $\Zop_N\times\Zop_N$ orbifold for even $N$}
\subsec{The theory without discrete torsion}
For $N$ even, the construction of the conventional fractional BPS
branes for which all three $s_i$ are even and of the conventional
fractional branes for which precisely one $s_i$ is odd, is exactly the
same as for odd $N$. What does change, however, is the situation for
the D(r;1,1,1) branes: a copy of such a brane on the covering space
can now be mapped to itself by some elements of the
$\Zop_2\times\Zop_2$ subgroup of $\Zop_N\times\Zop_N$. This opens up
the possibility of having some kind of fractional D(r;1,1,1)
brane. It will turn out that the situation depends on whether $N$ is 
divisible by four or not. In the following we shall write $N=2M$; the
situation will then depend on whether $M$ is even or odd.

As in the case $N=2$ we expect that the fractional D(r;1,1,1) brane
will have a moduli space with three different branches that are the
fixed planes of $g_1$, $g_2$ and $g_3=g_1 g_2$. For the first branch,
a boundary state can be given by
\eqn\pfNeven{\eqalign{|D(r;1,1,1);&\, {\bf y},{\bf a},\epsilon,
\epsilon'\rangle=\cr
\sum_{m,n=0}^{M-1}\Bigl\{
&|D(r;1,1,1);{\bf y},g_1^mg_2^n {\bf a}\rangle_{{\rm NS-NS;U}} 
+\epsilon |D(r;1,1,1);{\bf y},g_1^mg_2^n{\bf a}\rangle_{{\rm R-R;U}}
\cr 
+ \epsilon'\Bigl(
&
|D(r;1,1,1);{\bf y},g_1^mg_2^n{\bf a}\rangle_{{\rm NS-NS;T}_{g_1^M}}
+ \epsilon 
|D(r;1,1,1);{\bf y},g_1^mg_2^n{\bf a}
   \rangle_{{\rm R-R;T}_{g_1^M}}\Bigr)
\cr
+&  |D(r;1,1,1);{\bf y},g_1^mg_2^n({\bf - a})\rangle_{{\rm NS-NS;U}}  
+\epsilon 
|D(r;1,1,1);{\bf y},g_1^mg_2^n ({\bf - a})\rangle_{{\rm R-R;U}}
\cr 
- \epsilon'\Bigl(  &
|D(r;1,1,1);{\bf y},g_1^mg_2^n({\bf - a})\rangle_{{\rm NS-NS;T}_{g_1^M}}  
\cr
& \qquad \qquad
+ \epsilon 
|D(r;1,1,1);{\bf y},g_1^mg_2^n({\bf -a})
   \rangle_{{\rm R-R;T}_{g_1^M}}\Bigr)\Bigr\}
\,;}}
the construction for the other branches is analogous. We can read off
from this boundary state that the representation of the
$\Zop_2\times\Zop_2$ subgroup of $\Zop_N\times\Zop_N$ on the 
Chan-Paton indices is given by the direct sum of $M^2$ copies of
\repsorig\ (the multiplicity of $M^2$ is due to the fact that $M^2$ 
different orientations are necessary to make an orbifold invariant
configuration),
\eqn\Ztwosub{\eqalign{
\gamma(g_1^M)=&1_{M\times M}\otimes 1_{M\times M}\otimes 
\left( \matrix{1&0 \cr 0 & -1}\right)\cr
\gamma(g_2^M)=&1_{M\times M}\otimes 1_{M\times M}\otimes
\left( \matrix{0& 1 \cr 1 & 0}\right)\,, 
}}
where we have chosen a particular sign in \repsorig. This defines a
projective representation of $\Zop_2\times\Zop_2$, 
\eqn\projreptwo{
\gamma(g_1^M)\gamma(g_2^M)=-\gamma(g_2^M)\gamma(g_1^M)
}
which is consistent with the projective $\Zop_2\times\Zop_2$ action 
\eqn\commbis{ g_1^M g_2^M = - g_2^M g_1^M }
on the oscillator states of an open string between this brane and a
conventional fractional D(r';0,0,0) brane (see the discussion for
$N=2$ in Section~2). 

Of course, we need to study the action of the whole
$\Zop_N\times\Zop_N$ orbifold group on the open string states. From
the discussion for the $\Zop_2\times\Zop_2$ subgroup, it might seem
natural to expect a projective representation of $\Zop_N\times\Zop_N$
on the oscillator states and on the Chan-Paton factors. However, as we
shall now show, this can only be the case if $M$ is odd, \ie\ if $N$
is not divisible by four. In order to see this, let us recall that a
projective representation $r$ satisfies 
\eqn\projrep{
r(g_1)r(g_2)=\widehat\omega^{-1}r(g_2)r(g_1)\,.}  
Since $g_1^N=e$, it follows from \projrep\ that 
\eqn\omegaN{\widehat\omega^N= \widehat\omega^{2M}=1\,.}
Similarly, we can derive from \projrep\ that
\eqn\projreptwobis{
r(g_1^M)r(g_2^M)
=\widehat\omega^{-M^2}r(g_2^M)r(g_1^M)\,,}
so that consistency with \commbis\ or \projreptwo\ would require
\eqn\omegaMsquared{\widehat\omega^{M^2}=-1\,.}
It is then clear that \omegaN\ and \omegaMsquared\ are only consistent
if $M$ is odd. Let us therefore study odd and even $M$ separately. 
\smallskip

For odd $M$, we can define the action of the orbifold generators on
the $MN\times MN$ Chan-Paton factors to be given by
\eqn\projCP{\eqalign{
\gamma(g_1)=&\pmatrix{0&0&\cdots&0&1
\cr
1&0&\cdots&0&0
\cr
0&1&\cdots&0&0
\cr
\vdots&\vdots&&\vdots&\vdots
\cr
0&0&\cdots&1&0}_{M\times M}
\otimes 
{\rm diag}(1,\widehat\omega^{-1},\ldots,\widehat\omega^{-(N-1)})_{N\times N}
\cr 
\gamma(g_2)=&{\rm diag}(1,1,\ldots,1)_{M\times M}\otimes
\pmatrix{0&0&\cdots&0&1
\cr
1&0&\cdots&0&0
\cr
0&1&\cdots&0&0
\cr
\vdots&\vdots&&\vdots&\vdots
\cr
0&0&\cdots&1&0}_{N\times N}\,,
}}
where $\widehat\omega^M=-1$. This defines a projective representation of
$\Zop_N\times\Zop_N$,
\eqn\projrepZNZN{
\gamma(g_1)\gamma(g_2)=\widehat\omega^{-1}\gamma(g_2)\gamma(g_1)\,,
}
and it defines a representation equivalent to \Ztwosub\ for the
$\Zop_2\times\Zop_2$ subgroup generated by $g_1^M$ and $g_2^M$. This
is sufficient to guarantee that the resulting open string satisfies
the only easily testable consistency condition which comes from the
action of $g_1^M$ and $g_2^M$. 
\smallskip

For even $M$, \ie\ for $N$ a multiple of four, it follows from \omegaN\
and \omegaMsquared\ that the action of $g_1$ and $g_2$ on the
oscillator states of the open string does not even define a
representation that can be written as a direct sum of conventional and
projective representations. (The action of $g_1$ and $g_2$ in the
sector that describes the open strings between the D(r;0,0,0) and the
D(r';1,1,1) brane is not even a projective representation.) In order
for the total action to be consistent, this then implies that the same
has to be the case for the action on the Chan-Paton indices. It is not
difficult to see that one can choose an action on the Chan-Paton
indices that reproduces \projreptwo, and that has the property
\eqn\reppp{ \gamma(g_1) \gamma(g_2) 
    = {\cal R} \gamma(g_2) \gamma(g_1) \,,} 
where ${\cal R}$ is a diagonal matrix whose diagonal elements are
$\pm 1$. (If $\gamma$ was a projective representation, 
${\cal R}\propto {\bf 1}$; the above is therefore a mild
generalisation of a projective representation.) For example we can
define the $MN\times MN$ matrix  
\eqn\realise{
\gamma(g_1)=\pmatrix{ 
P & 0 & 0 & \cdots & 0 & 0 & 0 \cr
0 & P & 0 & \cdots & 0 & 0 & 0 \cr
0 & 0 & P & \cdots & 0 & 0 & 0 \cr
\vdots & \vdots & \vdots &  \vdots & \vdots & \vdots & \vdots \cr
0 & 0 & 0 & \cdots & \widehat{P} & 0 & 0 \cr
0 & 0 & 0 & \cdots & 0 & \widehat{P} & 0 \cr
0 & 0 & 0 & \cdots & 0 & 0 & \widehat{P} }\,,}
where the first $M$ matrices on the diagonal are 
\eqn\Pmat{ 
P = \pmatrix{0&0&\cdots&0&1
\cr
1&0&\cdots&0&0
\cr
0&1&\cdots&0&0
\cr
\vdots&\vdots&&\vdots&\vdots
\cr
0&0&\cdots&1&0}_{M\times M} \,,}
while the second $M$ matrices on the diagonal are 
\eqn\Phatmat{ 
\widehat{P} = \pmatrix{0&0&\cdots&0&-1
\cr
1&0&\cdots&0&0
\cr
0&1&\cdots&0&0
\cr
\vdots&\vdots&&\vdots&\vdots
\cr
0&0&\cdots&1&0}_{M\times M} \,.}
We also define $\gamma(g_2)$ to be the $MN\times MN$ matrix
\eqn\gammatwo{
\gamma(g_2)= \pmatrix{0&0&\cdots&0&{\bf 1}_M
\cr
{\bf 1}_M&0&\cdots&0&0
\cr
0&{\bf 1}_M&\cdots&0&0
\cr
\vdots&\vdots&&\vdots&\vdots
\cr
0&0&\cdots&{\bf 1}_M&0}\,,} 
and it is then not difficult to check that these matrices obey
\projreptwo, and define a `representation' satisfying \reppp\ where
${\cal R}$ has $\pm 1$ on the diagonal. This action on Chan-Paton
factors can be combined with an action on the oscillator ground states
of the open strings to give a conventional total action on the open
string space of states. For example, for the open string between a
fractional D(r;1,1,1) brane and a conventional fractional D(r';0,0,0)
brane, we can choose the action on the ground states of the different
open string sectors to be given by the above matrices. Then the total
action in \total\ is a conventional representation of the orbifold
group. The action on the
open string ground states so defined is consistent with the canonical
action defined for $g_1^M$ and $g_2^M$ in terms of fermionic zero
modes, as follows from the fact that the above matrices reproduce
correctly the commutation relations for $g_1^M$ and $g_2^M$.


\subsec{The theory with minimal discrete torsion}

As in the theory without discrete torsion, the only real difference
with the analysis for odd $N$ concerns the D(r;1,1,1) brane. Again,
the analysis depends on whether $N=2M$ is a multiple of four, and we
shall therefore consider the two cases ($M$ odd and $M$ even)
separately.

For $M$ odd, one may expect that the situation is quite similar to the
case $M=1$ ($N=2$) that we discussed in Section~2, and this is indeed
true. The relevant boundary state is of the form 
\eqn\pfModdDT{\eqalign{|D(r;1,1,1);&\, {\bf y},\epsilon,
\epsilon_1,\epsilon_2\rangle=\cr
\sum_{m,n=0}^{M-1} g_1^mg_2^n 
\Bigl\{&|D(r;1,1,1);{\bf y}\rangle_{{\rm NS-NS;U}} 
+\epsilon |D(r;1,1,1);{\bf y}\rangle_{{\rm R-R;U}}  \cr
& + \epsilon_1 \Bigl(
 |D(r;1,1,1);{\bf y}\rangle_{{\rm NS-NS;T}_{g_1^M}}
+ 
\epsilon |D(r;1,1,1);{\bf y}\rangle_{{\rm R-R;T}_{g_1^M}} \Bigr)
\cr
& + \epsilon_2 \Bigl(
 |D(r;1,1,1);{\bf y}\rangle_{{\rm NS-NS;T}_{g_2^M}}
+ 
\epsilon |D(r;1,1,1);{\bf y}\rangle_{{\rm R-R;T}_{g_2^M}} \Bigr)
\cr
& + \epsilon_1 \epsilon_2 \Bigl(
 |D(r;1,1,1);{\bf y}\rangle_{{\rm NS-NS;T}_{(g_1g_2)^M}}
+ 
\epsilon |D(r;1,1,1);{\bf y}\rangle_{{\rm R-R;T}_{(g_1g_2)^M}} \Bigr) 
\Bigr\}\,,}}
where the action of the group elements includes discrete torsion phases.
Since we are considering the case of minimal discrete torsion,
$\omega^{M^2}=-1$, each of the boundary states is invariant under
$g_1^M$ and $g_2^M$. Together with the fact that we have summed
together $M^2$ copies, this implies that the resulting boundary state is 
invariant
under the action of the orbifold group. It is not difficult to see
that it gives rise to a conventional action of the orbifold group on
the Chan-Paton indices. The consistency of this conventional
fractional D(r;1,1,1) brane can be tested as before, and the analysis is
again analogous to the case $N=2$.
\smallskip

If $M$ is even, \ie\ if $N$ is a multiple of four, $\omega^{M^2}=+1$,
and therefore the boundary components of \pfModdDT\ in the twisted
sectors are not invariant under the action of the orbifold group. In
order to construct a non-trivial boundary state we therefore have to
consider $NM(=2M^2)$ copies of the boundary states (rather than
$M^2$); the relevant boundary state is then given as 
\eqn\pfNevenDT{\eqalign{|&D(r;1,1,1);\, {\bf y},{\bf a},\epsilon,
\epsilon'\rangle=\cr
& \sum_{m,n=0}^{M-1}\Bigl\{
|D(r;1,1,1);{\bf y},g_1^mg_2^n{\bf a}\rangle_{{\rm NS-NS;U}} 
+\epsilon 
|D(r;1,1,1);{\bf y},g_1^mg_2^n{\bf a}\rangle_{{\rm R-R;U}}  \cr
& \quad + \epsilon' \omega^{-Mn} \Bigl(
|D(r;1,1,1);{\bf y},g_1^mg_2^n{\bf a}\rangle_{{\rm NS-NS;T}_{g_1^M}}
+ \epsilon 
|D(r;1,1,1);{\bf y},g_1^mg_2^n{\bf a}\rangle_{{\rm R-R;T}_{g_1^M}}
\Bigr) \cr
& + |D(r;1,1,1);{\bf y},g_1^mg_2^n({\bf - a})\rangle_{{\rm NS-NS;U}}  
+\epsilon 
|D(r;1,1,1);{\bf y},g_1^mg_2^n({\bf - a})\rangle_{{\rm R-R;U}} \cr 
& \quad - \epsilon'\omega^{-Mn} \Bigl(
|D(r;1,1,1);{\bf y},g_1^mg_2^n({\bf - a})\rangle_{{\rm NS-NS;T}_{g_1^M}}  
\cr
& \qquad \qquad \qquad \qquad 
+ \epsilon 
|D(r;1,1,1);{\bf y},g_1^mg_2^n({\bf -a})
\rangle_{{\rm R-R;T}_{g_1^M}}\Bigr) \Bigr\}
\,.}}
It is not difficult to check that this boundary state is invariant
under the action of the orbifold group. 

As before for the case without discrete torsion, the action on the
Chan-Paton factors that follows from this boundary state does not
even define a projective representation, but only satisfies \reppp. In
fact we have again that 
$\gamma(g_1^M) \gamma(g_2^M)= - \gamma(g_2^M)\gamma(g_1^M)$,
and therefore this is necessarily so. However, it can be checked as in
the theory without discrete torsion (and with similar limitations)
that this brane is consistent with the other branes of the theory. For
example, the open string to the D(r';0,0,0) brane is consistent since
the above sign is cancelled by the sign appearing in \comm, and the
action of $g_1^M$ and $g_2^M$ on the Chan-Paton indices of the
D(r';0,0,0) brane commute since $g_1$ and $g_2$ act by a projective
representation (with $\omega^{M^2}=+1$). The other cases are similar.

\newsec{Conclusions}

In this paper we have analysed the D-brane spectrum of orbifold
theories with and without discrete torsion. We have carefully examined
the consistency of the orbifold action on the open string states that
correspond to the different open strings between the various D-branes. 
Already for the simplest interesting example,
$\Cop^3/\Zop_2\times\Zop_2$, we have found that the analysis falls
outside the scope of the framework discussed in \gimpol; in fact, the
consistency requires that the D-brane spectrum contains branes that
carry a conventional representation of the orbifold group as well as
branes for which the representation is projective. This is precisely
in agreement with the D-brane spectrum that had been proposed in
\refs{\gab} using the boundary state formalism and the constraints of
T-duality. We have also analysed the non-BPS branes for this theory,
for which additional subtleties arise. In particular, the consistency
of the various symmetry operators requires that $(-1)^F$ acts
non-trivially on the Chan-Paton indices of some non-BPS branes. As
before, this is beautifully reproduced by the corresponding boundary
states that we construct. 

We have also analysed how these results generalise to the orbifolds 
$\Cop^3/\Zop_N\times\Zop_N$. Among other things we have found that
some of these theories have non-BPS D-branes carrying
R-R charges that are not carried by any of the (known) BPS branes of
the theory. These non-BPS D-branes enjoy special stability
properties. 

Most of our analysis has been done case by case, and it would be
interesting to be able to understand these results more conceptually.
In particular, it should be possible to understand the nature of the
representation of the orbifold group on the Chan-Paton indices of a
given brane more abstractly, for example in terms of K-theory. The
consistency of the various open string actions should then also follow 
from some abstract arguments. 

\vskip 1cm

\centerline{{\bf Acknowledgements}}\pano

We would like to thank Eric Gimon, Jeff Harvey, Emil Martinec, George
Papadopoulos, Fred Roose and Ashoke Sen for useful conversations. 

The work of BC is supported by DOE grant DE-FG02-90ER-40560 and NSF grant  
PHY-9901194. MRG is grateful to the Royal Society for a University
Research Fellowship. He also acknowledges partial support from the EU
networks ``On Integrability, Nonperturbative effects, and Symmetry in
Quantum Field Theory'' (FMRX-CT96-0012) and ``Superstrings''
(HPRN-CT-2000-00122), as well as from the PPARC special grant
``String Theory and Realistic Field Theory'', PPA/G/S/1998/0061.

\listrefs

\bye